\documentclass{article}

\usepackage{arxiv}

\usepackage[utf8]{inputenc} 
\usepackage[T1]{fontenc}    
\usepackage{hyperref}       
\usepackage{url}            
\usepackage{booktabs}       
\usepackage{amsfonts}       
\usepackage{nicefrac}       
\usepackage{color,xcolor}
\usepackage{microtype}      
\usepackage{graphicx}
\usepackage{amssymb}
\usepackage{amsmath}

\newcommand{\textblue}[1]{\textcolor{black}{{#1}}}

\title{Topological data analysis approaches to uncovering the timing of ring structure onset in filamentous networks}

\author{
  Maria-Veronica Ciocanel\\
  Duke University\\
  \texttt{ciocanel@math.duke.edu} \\
   \And
 Riley Juenemann \\
  Department of Mathematics\\
  Tulane University\\
  \And
 Adriana T.~Dawes \\
   Department of Mathematics\\
   Department of Molecular Genetics\\
   The Ohio State University\\
  \And
 Scott A.~McKinley \\
  Department of Mathematics\\
  Tulane University\\
}

\begin{document}
\maketitle

\begin{abstract}
In developmental biology as well as in other biological systems, emerging structure and organization can be captured using time-series data of protein locations. In analyzing this time-dependent data, it is a common challenge not only to determine whether topological features emerge, but also to identify the timing of their formation. For instance, in most cells, actin filaments interact with myosin motor proteins and organize into polymer networks and higher-order structures. Ring channels are examples of such structures that maintain constant diameters over time and play key roles in processes such as cell division, development, and wound healing. Given the limitations in studying interactions of actin with myosin \textit{in vivo}, we generate time-series data of protein polymer interactions in cells using complex agent-based models. Since the data has a filamentous structure, we propose sampling along the actin filaments and analyzing the topological structure of the resulting point cloud at each time. Building on existing tools from persistent homology, we develop a topological data analysis (TDA) method that assesses effective ring generation in this dynamic data. This method connects topological features through time in a path that corresponds to emergence of organization in the data. In this work, we also propose methods for assessing whether the topological features of interest are significant and thus whether they contribute to the formation of an emerging hole (ring channel) in the simulated protein interactions. In particular, we use the MEDYAN simulation platform to show that this technique can distinguish between the actin cytoskeleton organization resulting from distinct motor protein binding parameters.  
\end{abstract}


\section{Introduction}
Topological data analysis (TDA) \textblue{has emerged as a new and important set of statistical tools for extracting structural information from high-dimensional data sets} \cite{edelsbrunner2002topological,edelsbrunner2008persistent,edelsbrunner2010computational}. In particular, methods from persistent homology are useful in understanding topological invariants such as clusters or loops in data represented as point clouds. Applications of these topological methods \textblue{in the biological sciences} are varied and include quantitative understanding of aggregations such as insect swarms \cite{topaz2015topological,ulmer2019topological}, extracting the topology of functional brain networks from fMRI data \cite{saggar2018towards,stolz2018topological}, and mapping of unknown spatial environments using biobotic insects \cite{dirafzoon2016geometric}. \textblue{While the applications for TDA multiply, there is an associated need for statistical methods that can rigorously compare the behaviors of systems when they are subjected to different environments or controls \cite{wasserman2018topological}. A particular challenge for TDA methods is that complex data sets are often corrupted by noisy observations and/or missing data \cite{chazal2017robust,bobrowski2017topological}. Moreover, the methods of TDA often exhibit spurious topological features that can vastly outnumber the ``real'' topological features of interest \cite{fasy2014confidence}. This last issue is a particular challenge for the biological systems that motivate the work we present here.}  

\subsection{\textblue{Biological context: ring formation in filamentous networks}}
\textblue{There are numerous active scientific questions concerning the formation and maintenance of ring-like actin filament structures,} which play key roles in developmental and physiological processes \cite{schwayer2016actin}. 
These ring channels are usually composed of actin filaments cross-linked with myosin motor proteins as well as other regulatory binding proteins that control the spatiotemporal organization of the filaments into circular structures in living systems \cite{schwayer2016actin}. 
These actin-myosin rings have been shown to participate in both actively contracting rings, such as those found in cytokinesis and wound healing in organisms ranging from plants to mammals \cite{robinson1996stable,schwayer2016actin}, as well as in stable ring-like structures that are often used as inter-cellular bridges by germline cells to share nutrients and gene products during development \cite{robinson1996stable}.

We focus on stable biological ring-like structures that maintain a constant diameter over a long time. Two specific examples of such stable structures occur in developing germline cells in \textit{Drosophila} fruit flies and in \textit{Caenorhabditis} nematode worms.
In \textit{Drosophila} fruit fly development, ovarian ring canals connect germ cells and promote transport of cellular components to the developing egg \cite{robinson1994morphogenesis,hudson2015actin}. The complex assembly and maintenance of these ring canals is not fully understood and there is evidence that various specialized cytoskeletal proteins may regulate their development \cite{robinson1994morphogenesis,ong2010mutations,hudson2015actin}. 
Similarly, the actin cytoskeleton forms complex cellular structures in the reproductive system of the worm \textit{C. elegans} \cite{kelley2019regulation}. Here, actin filaments interact with non-muscle myosin II motor proteins and other actin-binding proteins to allow for streaming of cytoplasm into enlarging oocytes \cite{wolke2007actin,osorio2018flow}. The stable circular structures that emerge are called ring channels and maintain a constant diameter during the development of germ cells into oocytes \cite{coffman2016antagonistic}. However, the timing of onset of these ring channels and the mechanisms that contribute to their maintenance through time are not well understood.

Given the challenges involved in visualizing actin dynamics in these \textit{in vivo} systems, complex simulations of actin-myosin networks as proposed in \cite{popov2016medyan} provide useful tools for studying the dynamics and remodeling of these cellular structures. Filament contractility and alignment in these agent-based modeling simulations have been assessed by calculating the network radius of gyration and the orientational order parameter for all actin \textblue{monomer units (}segments\textblue{)} simulated \cite{popov2016medyan}. However, questions related to the timing and maintenance of ring channel formation as well as to their regulation by cytoskeletal proteins remain unanswered.

\subsection{\textblue{Topological data analysis for time-dependent data}}
\textblue{We investigate simulations of actin-myosin interactions where we qualitatively observe the emergence of one global hole in the simulation domain. When extracting topological information from this data, we are therefore interested in the most significant $1$-dimensional hole corresponding to a ring channel (rather than other noisy features). Tracking topological features and their evolution through time is a natural question in this context; several studies have addressed aspects of this question, but not for our setting. For example,} the crocker (Contour Realization of Computed k-dimensional hole Evolution in the Rips Complex) plots developed in \cite{topaz2015topological} investigate time dynamics using topological data analysis.  \textblue{Crocker plots} keep track of Betti numbers of point clouds generated by dynamical systems models of biological aggregations. The authors show that their proposed quantitative and visualization tools have predictive value in selecting models describing agent interactions \cite{ulmer2019topological}. \textblue{While these visual tools have been an inspiration for this work, we are interested in tracking the birth and death scale for a significant hole (ring) over consecutive time frames of the actin-myosin simulations, rather than reporting the number of topological holes with time. Our method therefore focuses on connecting features through time using topological summaries called persistence diagrams.} 

\textblue{A much greater theoretical challenge is to assess the continuity of a topological object through time in the original image space. Several studies  addressing this challenge have been proposed and adapted to data in other biological systems.} The approach in \cite{cohen2006vines} introduces vineyards, which are time-parameterized stacks of persistence diagrams requiring computation of simplices \textblue{(using sublevel sets filtration)} at each time point. These methods are applied to protein folding trajectories in \cite{cohen2006vines}. The study of \cite{kim2020analysis} views dynamic data sets as time-varying graphs and extracts summaries of their clustering features, with potential applications to swarming behaviors. \textblue{Our work is an approximation of such theoretical approaches that is accurate for extracting the most significant feature and its emergence through time. This approach generates a path (corresponding to a vine in \cite{cohen2006vines}) using a computationally efficient algorithm based entirely on topological summaries of the data (persistence diagrams).} 

\textblue{Since the actin-myosin simulations studied here consistently show the emergence of a global hole, our method for tracking a significant topological feature through time is complementary to this observation.} This method provides an intuitive computational approach to connecting through time significant features \textblue{within a} persistence diagram corresponding to persistent topological holes. \textblue{The resulting time-dependent path allows us to determine whether and when a structure corresponding to a ring channel emerges. Our setting applies to} data represented as point clouds naturally or extracted \textblue{by sampling} from filamentous networks. \textblue{In this work, we address how the density of sub-sampling of points from filaments affects conclusions related to the timing of ring formation in polymer networks.} \textblue{The proposed technique} provides insights into the timing of higher-order structure formation and organization in time series data \textblue{representing actin-filament interactions.}

\subsection{\textblue{Statistical significance in topological data analysis}}
\textblue{This work raises the natural question of when topological features become statistically significant, which is an outstanding problem in statistical topological data analysis. One of the most prominent examples of this kind of work is a series of papers by Fasy, Chazal, and others \cite{chazal2013bootstrap,fasy2014confidence,chazal2017robust, bobrowski2017topological}. In these works, the authors generally assume that there is a static compact set $S$ on which a probability distribution $P$ is supported. The goal is to infer the topological structure of $S$ by analyzing a finite number of samples from $P$. \cite{fasy2014confidence} established a notion of confidence sets for persistence diagrams -- the idea that the true persistence diagram will fall within a threshold distance of an observed persistence diagram, say, $95\%$ of the time. In establishing these confidence sets, the authors observe that when the distance between the true persistence diagram and the observed one is sufficiently small, then any topological features with persistence less than the threshold value cannot be a feature of $S$. These features are therefore considered noise that arises from the finite-size sampling of $P$. Throughout this work, we will call such features ``spurious,'' ''not significant,'' or simply ``noisy'' depending on the context.}

\textblue{The notion of confidence sets requires that there is a ``true'' underlying persistence diagram that is to be estimated. Our challenge is slightly different in that the topological features are emergent from a dynamic filament network and the structures are fully qualitative.  That is to say, there is no actual hole in the network, but often the appearance of a hole is unmistakable to the eye. Moreover, the structure of the noise underlying spurious topological features in our simulations is different from that assumed in much of the theoretical literature. In previous studies, the spurious features either arise from the discrete sampling from $P$ or from a combination of this sampling and some model for outlier samples (with an explicit outlier probability distribution) \cite{chazal2017robust}. In our case, the noise comes from filaments that are not a part of the emergent structure and their distribution is difficult to model.} 

\textblue{There are yet more perspectives on statistical aspects topological data analysis. For example, there is work that does not use persistence diagrams as a central summary object. Notably,  \cite{bubenik2015statistical} proposed the notion of persistence landscapes in order to create a summary that is well-defined in a Hilbert space, leading then to a notion of $p$-values in that space. \cite{blumberg2014robust} studied distributions of barcodes as a basis for understanding hypothesis testing and confidence sets. Recently, \cite{maroulas2020bayesian} retained the persistence diagrams as the summary tool, but took a Bayesian perspective on inference.}

\textblue{In this work, the topological structure is simple, in the sense that we are looking for at most one hole; however, the analysis is difficult due to the filamentous nature of the point cloud and the unusual structure of noise that arises from the stochastic dynamics. As a result, the simplicity of structure results in visual summaries that tell a very clear story; but, the complexity of the noise interferes with a direct application of existing theory. In light of these considerations, we present three different methods for addressing the question of what constitutes \emph{significance} of a topological feature in these simulations. In the end, these different perspectives roughly align in terms of the threshold for significance that they produce.}

\textblue{First, we take a hypothesis testing perspective, which requires the establishment of an appropriate null model for randomly distributed filament networks. Using computational methods, we essentially employ a filamentous version of the Poisson spatial process null hypothesis that has been studied by \cite{bobrowski2017maximally} and \cite{bobrowski2018topology}. 
In a second approach, we directly study the distribution of topological features that arise from simulations where filaments interact with molecular motors. The empirical cumulative distribution function has a plateau over an interval of persistence lengths that can be used to qualitatively infer a distinction between significant and spurious features. Finally, in a third perspective, rather than considering the simulations frame-by-frame (and pooling across frames when appropriate), we study the time-dependent paths themselves to assess significance. We do this by calculating the persistence diagram trajectory (approximating the vine concept developed in \cite{cohen2006vines}) associated with the topological feature that has maximal persistence across an entire simulation and comparing it to frame-by-frame maximally persistent features. The persistence level at the emergence time provides an estimate for significance that is similar to the first two methods.}

\subsection{Overview}

\textblue{The manuscript is organized as follows. In \S~\ref{sec:medyan}, we describe the MEDYAN simulation framework for actin-myosin interactions \cite{popov2016medyan} and give an overview of the simulation databases generated and used in this work. In \S~\ref{sec:tda}, we provide a review of the key topological data analysis concepts that we use, including the persistence diagram summary. We discuss the algorithm for tracking topological features through time in \S~\ref{sec:tda_time} and illustrate its application to a sample simulation. In \S~\ref{sec:significance}, we describe three methods for determining a significance threshold in persistence diagrams and show their agreement for time-series data from our actin-myosin simulation databases. In \S~\ref{sec:analysis}, we show how the proposed methods distinguish between and provide insights to simulations of actin-myosin interactions with different motor binding parameters. We conclude with a short discussion in \S~\ref{sec:conclusions}.
}

\section{Methods}
\label{sec:methods}
\subsection{Stochastic simulation framework for actin-myosin interactions}\label{sec:medyan}

\subsubsection{\textblue{MEDYAN simulation framework}}
\label{sec:medyan_general}

For simulations of actin-myosin interactions, we use the MEDYAN model developed by the Papoian Lab and introduced in \cite{popov2016medyan}. This agent-based modeling framework simulates actin filaments as interacting semi-flexible polymers in a solution with complex reaction and diffusion processes in three dimensions. The actin filaments interact with motor proteins such as myosins and with transient cross-linking proteins. The numerical method involves simulating a three-dimensional stochastic reaction-diffusion scheme for the active matter model using a spatially resolved Gillespie algorithm.

MEDYAN models chemical phenomena on a simulation space that is divided into compartments. Diffusion and molecular transport of various chemical species are modeled as stochastic jumps between compartments. For the purposes of our simulations, these dynamics include growth and shrinking of actin filaments, cross-linker and molecular motor binding, and active transport by molecular motors (walking). The model also uses a mechanical representation of the actin filament network where the filaments consist of multiple cylindrical \textblue{segments (which we will refer to as \textit{monomer units})} that simulate semi-flexible polymers with a given persistence length. The model includes various interaction potentials for filament deformations as they interact with other structures in the simulation domain. Additional information on details of the MEDYAN model and implementation can be found in \cite{popov2016medyan,komianos2018stochastic}. 
In \textblue{most} simulations \textblue{and unless otherwise noted}, we use a standard implementation of the model in \cite{popov2016medyan}, which is parameterized for an actin-myosin network consisting of actin filaments, $\alpha$-actinin cross-linking proteins, and non-muscle myosin IIa motor filaments.

\subsubsection{\textblue{MEDYAN simulations with varied motors and linkers}}
\label{sec:medyan_database}

\textblue{In order to develop an overview of actin-myosin interactions that reliably develop ring structures, we generated a collection of 35 MEDYAN simulations (with 200 time frames each)  that have a fixed motor parameter set, but different numbers of linkers and motors; there are many motor parameters in the MEDYAN model framework, therefore we use the standard myosin-2 parameters that can be found in \cite{popov2016medyan}. In our simulations, the motor numbers range from 0 to 10, while the linker numbers range from 0 to 3000 (these ranges include the standard values in \cite{popov2016medyan}). We used this collection to study the distribution of persistence lengths discussed in \S~\ref{sec:significance}}.

\subsubsection{\textblue{MEDYAN simulations with varied binding rates}}
\label{sec:medyan_small_large}

\textblue{In order to assess the impact of motor binding rates on ring formation, we generated two more collections of simulations. The parameter we vary (which we refer to as the on-rate) is the rate constant of the binding reaction linking a myosin motor to two actin filaments. In the standard MEDYAN simulations in \cite{popov2016medyan}, this on-rate has a value of $0.2$~s$^{-1}$. We generate 40 simulation runs with an increased on-rate ($0.4$~s$^{-1}$) and 40 runs with decreased on-rate ($0.1$~s$^{-1}$). Qualitatively, we observed that the small on-rate reliably produced a ring structure in the experimental window, but the high on-rate did not. These are the collections of simulations that we study in \S~\ref{sec:analysis} when we put our proposed methods into practice.}

\subsection{Topological data analysis}\label{sec:tda}
We analyze data from simulations of actin-myosin interactions using tools from topological data analysis. We give a brief introduction to the ideas behind these tools, but we refer the reader to \cite{topaz2015topological} for a nontechnical overview or \cite{ulmer2019topological} for a more detailed technical explanation of these techniques.

\subsubsection{Persistent homology}\label{sec:pershomology}
The data in our study consists of points in three-dimensional space that are extracted from the actin filaments in our simulations. Specifically, these data points correspond to $x$, $y$, and $z$ locations of the cylindrical monomers that make up the actin polymers. We consider this finite set of points as a sampling from the ambient space $\mathbb{R}^3$. Given this discrete set of points $S$, we build the Vietoris-Rips simplicial complexes. In our application, the distance between points is simply the Euclidean distance between the three-dimensional locations of the points (which are viewed as vertices or nodes). For each real-valued parameter $\epsilon>0$ (called proximity parameter), the Vietoris-Rips construction forms a $k$-simplex whenever $k+1$ points are pairwise within the distance $\epsilon$. For example, a 1-simplex (an edge) forms when two points are within distance $\epsilon$ of each another, and a 2-simplex forms when any pair of points from a set of three points are within distance $\epsilon$ of each another. There are several simplicial complex constructions that one may choose for point clouds extracted from data \cite{otter2017roadmap}. As in \cite{topaz2015topological}, we choose to use the Vietoris-Rips complex given that it is more computationally tractable than the corresponding \u{C}ech complex \cite{ghrist2008barcodes}. While the \u{C}ech complex provides a simplicial complex model that is faithful to the topology of the starting point cloud, the Vietoris-Rips complex approximates the \u{C}ech complex and, since it is a flag complex (maximal simplicial complex built from the underlying graph), it allows for efficient storage as a graph \cite{ghrist2008barcodes}.

\textblue{The Vietoris-Rips complex $S_\epsilon$ at scale $\epsilon > 0$ is defined as $S_\epsilon = \{\sigma \subseteq S \mid d(x,y) \le 2\epsilon \; \forall \, x,y \in \sigma \}$ \cite{otter2017roadmap}.} We are interested in using homology, a tool from algebraic topology, to count and record features such as connected components, holes, and voids (trapped volumes) in the resulting space $S_\epsilon$ of simplicial complexes. By imposing an algebraic structure on $S_\epsilon$, one can define the $k$th homology $H_k(S_\epsilon)$. The dimension of $H_k(S_\epsilon)$ is denoted as the Betti number $b_k$ and gives the number of $k$-dimensional holes. For example, the number of connected components is $b_0$, the number of topological holes is $b_1$, and so on. 

\textblue{For any given a value of the proximity parameter $\epsilon$, one can gain some insight for qualitative features of the data by building the associated simplicial complex $S_\epsilon$ and calculating its homology \cite{otter2017roadmap}. However, any single choice for $\epsilon$ can obscure key features. The key idea underlying persistent homology is to assess topology across all reasonable values of $\epsilon$.} As $\epsilon$ increases, simplices are added to the complexes, so that there is an inclusion of complexes for smaller $\epsilon$ into those arising from a larger $\epsilon$ scale. This gives a finite sequence of nested subcomplexes which forms a filtered simplicial complex. Persistent homology then records how the homology of the simplices changes as the proximity parameter increases, thus identifying features that persist across a range of $\epsilon$. In addition, persistent homology associates a lifetime interval to each feature, which corresponds to the range of parameters $\epsilon$ over which the feature persists. Here we focus on the persistence diagram visualization of persistent homology computations, which allows us to keep track of these lifetime intervals by tracking their birth and death coordinates. As we further explain later, Betti numbers only provide the total number of features at each scale $\epsilon$; the persistence diagram visualization allows us to focus on individual features and their persistence, therefore conveying useful information for our application. 

\begin{figure}[!h]
\caption{{\bf Point cloud from actin simulations and resulting simplicial complexes.}
(a) Sample simulation with five actin filaments in a $2000$~nm by $2000$~nm by $200$~nm domain. (b) The point cloud extracted from the simulation in (a), consisting of the center and ends of each filament. The points are projected in two dimensions for visualization (and not for analysis) purposes. \textblue{We later argue for more dense sampling of the filaments in obtaining the point clouds.} (c) The persistence diagram corresponding to the point cloud in (b). Black circles in the persistence diagram correspond to connected components, while red triangles correspond to $1$-dimensional holes. (d) The resulting Vietoris-Rips complexes of the $15$ points in the point cloud in (b) as the proximity parameter $\epsilon$ varies. Each point represents a 0-simplex, an edge is a 1-simplex if the $\epsilon/2$ circular neighborhoods of the two points intersect, a triangle is a 2-simplex if the vertices are pairwise connected by edges, etc. For ease of visualization, the right panel only shows the balls around six points in the domain and highlights in red a topological hole. Note that the complexes are visualized in two dimensions whereas the calculation in (c) and throughout the manuscript is carried out for the three-dimensional point clouds.}
\vspace{2ex}
\includegraphics[width=\textwidth]{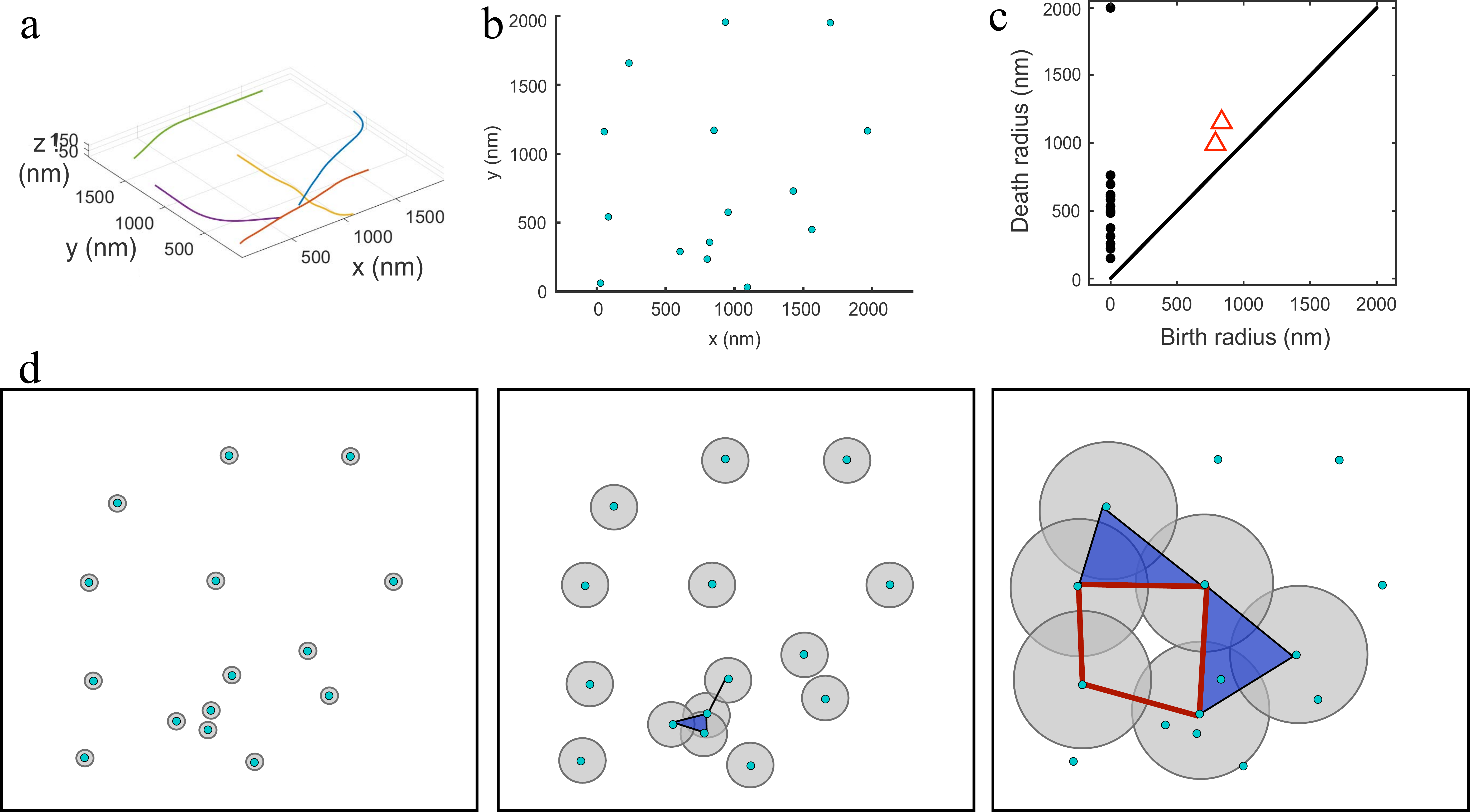}
\label{fig:simp_complex}
\end{figure}

Figure~\ref{fig:simp_complex}d shows an example of a finite filtered simplicial complex that forms as we vary the value of the proximity parameter $\epsilon$ for data points (Figure~\ref{fig:simp_complex}b) extracted from a simple MEDYAN simulation of actin polymer dynamics (Figure~\ref{fig:simp_complex}a). The point cloud represents a sampling of cylindrical \textblue{segment (monomer unit)} locations along the simulated filaments; for simplicity, we extract and visualize only the first, middle, and last \textblue{monomer unit} location for each filament in Figure~\ref{fig:simp_complex}. For all remaining simulations in this study, we sample $30\%$ of the \textblue{units} along each filament. In the Appendix, we illustrate the impact that sampling different percentages of the actin \textblue{monomers} in simulated filaments has on our proposed methods.

\subsubsection{Persistence diagrams and previous analysis of time-series data}\label{sec:PDs}

\begin{figure}[!h]
\caption{{\bf Time series data and persistence diagrams.}
(a) Sample time-series data of the actin-myosin interactions in a MEDYAN simulation; different actin filaments are depicted as long colorful polymers and cross-linkers are shown as short black lines. Myosin motors are omitted in this visualization for clarity; they are represented by medium-length dashed blue lines in the supporting videos (Online Resources 1 and 2). (b) Sampling of $30\%$ of the \textblue{monomer units} along the $50$ actin filaments in each time snapshot of the simulation. (c) Corresponding persistence diagrams generated by calculating the Rips filtration for each of the point clouds in (b); black circles correspond to connected components and red triangles correspond to loops.}
\vspace{2ex}
\includegraphics[width=\textwidth]{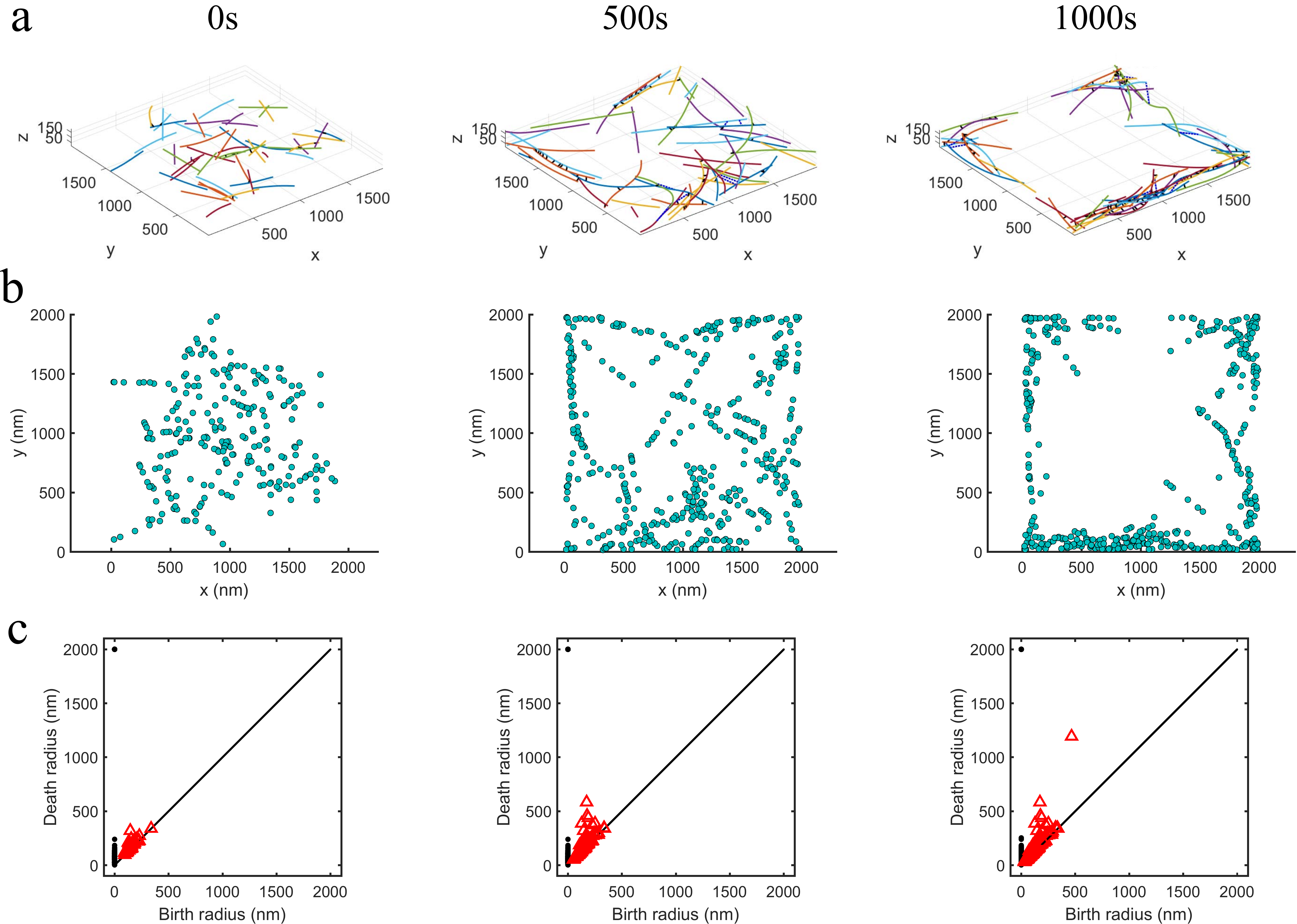}
\label{fig:pers_diagram_time}
\end{figure}

One common method introduced by \cite{edelsbrunner2002topological} for displaying information about the \textblue{persistent homology} of a set of data points is a persistence diagram, as shown in Figure~\ref{fig:simp_complex}c and in Figure~\ref{fig:pers_diagram_time}c. A persistence diagram is a multi-set of birth-death pairs corresponding to the birth and death of the homology generators. In other words, this visualization shows the scale $\epsilon$ at which a feature appears on the $x$-axis (birth radius) and the scale at which the feature disappears on the $y$-axis (death radius). We emphasize that while the terms ``birth" and ``death" connote time dependence, they in fact refer to the spatial scale over which a feature persists for a point cloud extracted from data at a fixed time point. Features that correspond to connected components ($0$-dimensional holes) always start at $\epsilon=0$. Features that correspond to loops ($1$-dimensional holes) are consistently above the main diagonal. Typically the most significant features are those corresponding to birth-death pairs that are farther away from the diagonal, because they persist over a wider range of scales, although \textblue{this is not always the case} \cite{feng2019persistent}. \textblue{In our work, we generate persistence diagrams} using the \texttt{ripsDiag} function in the \texttt{TDA} package in R, which calculates the Rips filtration built on top of a point cloud \cite{fasy2014introtda}. In particular, we use this function with the GUDHI C++ library for computing persistence diagrams \cite{maria2014gudhi}.

In many applications, it is informative to understand the topology of the data as it varies with time. The studies of \cite{topaz2015topological,ulmer2019topological} propose an approach to calculate and visualize the Betti numbers of dynamic data as a function of both the proximity parameter $\epsilon$ and time. Their method, called Contour Realization of Computed k-dimensional hole Evolution in the Rips Complex (crocker), keeps track of Betti numbers $b_k(\epsilon,t)$. The authors then use this matrix of data as feature vectors that help select appropriate models of biological aggregation for given experimental data in \cite{ulmer2019topological}. This approach of tracking the total numbers of topological features of each dimension across proximity scale and simulation time also helps identify group phenomena such as alignment and clustering \cite{topaz2015topological}.

\textblue{Our motivating scientific question is slightly different. Rather than being interested in the number of topological features, we are focused on identifying the emergence and duration of a single major feature.}
So, instead of counting the number of features at each time and persistence scale (as in crocker plots), we keep track of the birth and death coordinates in the persistence diagram corresponding to $1$-dimensional hole features at each time point. Recording both coordinates of these points (birth-death pairs) in the persistence diagram allows us to connect through time pairs that are close in this space and to identify statistically significant features and properties of a time-changing point cloud. \textblue{As discussed in the introduction, this goal is much closer to the concept of vines and vineyards developed by \cite{cohen2006vines} and more recently addressed by \cite{kim2020analysis}.}

\subsection{Topological data analysis for detecting rings in time-series data}\label{sec:tda_time}

Since we are interested in the formation of ring structures, we focus on the red triangles in the persistence diagrams of Figure~\ref{fig:pers_diagram_time}c. This figure shows three time frames in a simulation with $50$ actin filaments interacting with myosin motors and $\alpha$-actinin on a domain of realistic size for actin structure organization (see Figure~\ref{fig:pers_diagram_time}a). For the simulations in this study, we extract $30\%$ of the actin \textblue{monomer units} along each filament to generate the point cloud at each time step. In the Appendix, we discuss how coarser and finer samplings affect our results. We seek a method to determine the emergence of a significant topological hole, as illustrated by the red triangle farthest from the diagonal in the third column of Figure~\ref{fig:pers_diagram_time}c, corresponding to $1000$~s into the simulation. 

\subsubsection{Persistence of rings in time-series data}\label{sec:pers_rings}

\begin{figure}[!h]
\caption{{\bf Visualization of birth-death proximity parameter pairs with time in a dynamic simulation and the path generated using our method.}
(a) Black triangles denote the birth radii ($\epsilon_{\mathrm{birth}}$) corresponding to the formation of a topological hole, while red triangles denote the death radii ($\epsilon_{\mathrm{death}}$) corresponding to the disappearance of this structure. \textblue{(b) The displayed path is the output of our method: blue dots correspond to the significant feature's persistence ($\epsilon_\mathrm{death} - \epsilon_\mathrm{birth}$) as a function of time. Dashed lines correspond to the statistically-determined time of hole formation (vertical) and the statistically-determined significant distance from the diagonal (horizontal).}}
\vspace{1ex}
\centering
\includegraphics[width=0.9\textwidth]{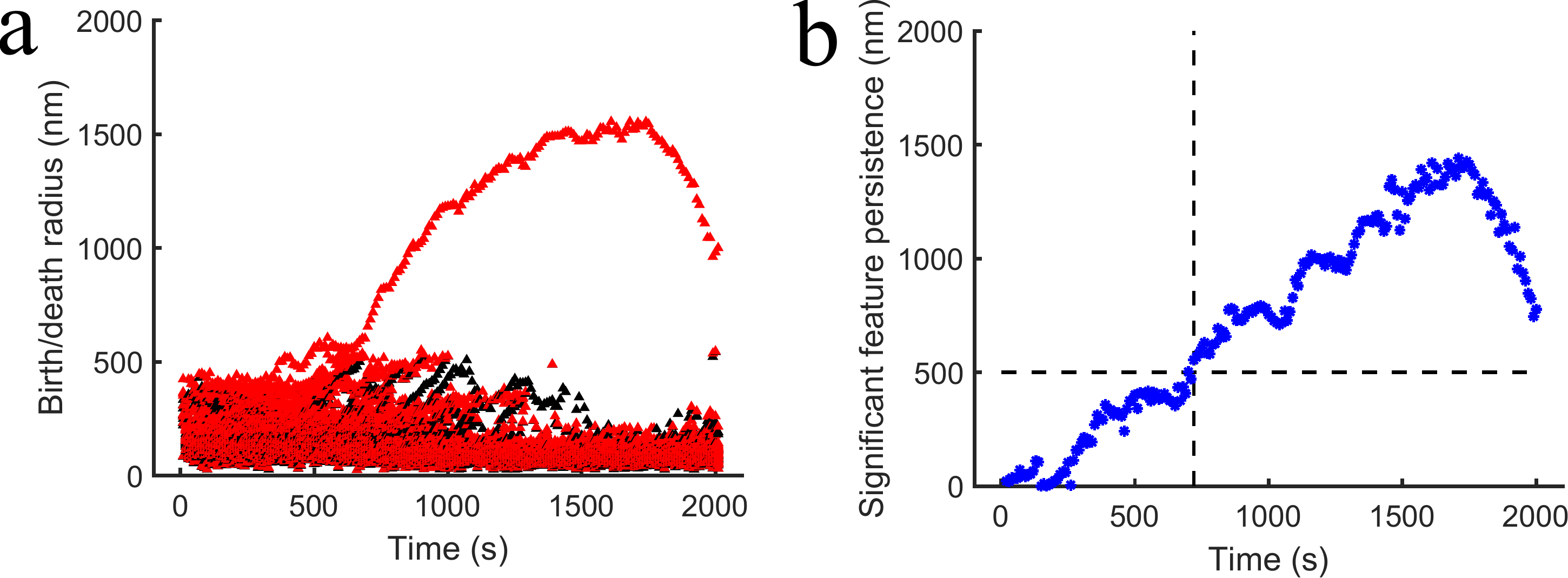}
\label{fig:bd_dvt}
\end{figure}

Persistence diagrams illustrate the birth radius $\epsilon_{\mathrm{birth}}$ on the $x$-axis and the death radius $\epsilon_{\mathrm{death}}$ on the $y$-axis corresponding to each topological feature, but do not allow for visualization of features across time. In order to track the evolution of these birth-death pairs in the persistence diagram over time, we constructed a visualization that overlays successive $(\epsilon_{\mathrm{birth}},\epsilon_{\mathrm{death}})$ pairs for all such features as they vary in time for the dynamical systems models of actin-myosin interactions considered. Figure~\ref{fig:bd_dvt}a shows an example of this visualization for the simulation in Figure~\ref{fig:pers_diagram_time}a, continued up to $2000$~s. The black triangles correspond to radii $\epsilon_{\mathrm{birth}}$ corresponding to the scale at which a $1$-dimensional hole arises, whereas the red triangles denote radii $\epsilon_{\mathrm{death}}$ corresponding to the scale at which the feature disappears. The hole that forms in the middle of the simulation domain is easily identified as the continuous evolution of a pair of $(\epsilon_{\mathrm{birth}},\epsilon_{\mathrm{death}})$ that diverges from other features, and remains an outlier. It is worth noting that most pairs of the birth and death proximity parameter likely amount to topological noise and are close to the diagonal in Figure~\ref{fig:pers_diagram_time}c, or have almost overlapping birth and death radii in Figure~\ref{fig:bd_dvt}a. \textblue{Figure~\ref{fig:bd_dvt}b shows the output of our method (generated as described in \S~\ref{sec:algorithm}): the extracted path through time for the most significant 1-dimensional feature in Figure~\ref{fig:bd_dvt}a. The displayed path corresponds to the difference between the specific pairs of death radii (red triangles in Figure~\ref{fig:bd_dvt}a) and birth radii (black triangles in Figure~\ref{fig:bd_dvt}a) for the path identified using our proposed algorithm.}

\subsubsection{Algorithm for visualization of ring structure persistence} \label{sec:algorithm}
To explore the emergence of these continuous ring structures in time-series data, we propose a method for connecting the pairs of $(\epsilon_{\mathrm{birth}},\epsilon_{\mathrm{death}})$ that are most likely to correspond to the same $1$-dimensional hole structure through time as illustrated by the pair emerging around time $600$~s in Figure~\ref{fig:bd_dvt}a. In the persistence diagram plots in Figure~\ref{fig:pers_diagram_time}c, our method will connect the red triangles corresponding to consecutive time points (at intervals of $10$~s). Our approach consists of the following steps:
\begin{enumerate}
\item Calculate all birth-death pairs $(\epsilon_{\mathrm{birth}},\epsilon_{\mathrm{death}})$ \textblue{corresponding to the dimension of interest} for each time step.
\item \textblue{At each time step, order the birth-death pairs from highest persistence to lowest persistence (i.e., order them in decreasing order of $\epsilon_{\mathrm{death}}-\epsilon_{\mathrm{birth}}$).}
\item Start by considering the first two time steps and calculate the matrix of \textblue{$L_\infty$} distances between all birth-death pairs at the first time step and the pairs at the next time step:
\begin{align}\label{eq:L_infty}
||(\epsilon_{\mathrm{birth1}},\epsilon_{\mathrm{death1}})&- (\epsilon_{\mathrm{birth2}},\epsilon_{\mathrm{death2}})||_\infty = \nonumber \\ & \max(|\epsilon_{\mathrm{birth2}}-\epsilon_{\mathrm{birth1}}|,|\epsilon_{\mathrm{death2}}-\epsilon_{\mathrm{death1}}|) \,. 
\end{align}
\item For each pair in the first time step \textblue{and starting with the most persistent feature}, find the pair in the second time step that is closest to it \textblue{in the sense of the $L_\infty$ metric in \eqref{eq:L_infty}}. If this smallest distance is greater than a parameter that we denote as the linkage tolerance $d_\ell$, then the pair is not counted as a connection.
\item Repeat steps 2-\textblue{4} for all consecutive time steps until the end of the simulation. Pairs at consecutive time points are then combined into paths (see Figure~\ref{fig:paths} \textblue{in the Appendix}).
\item Isolate the most significant path by finding the pair that is farthest from the diagonal ($\epsilon_{\mathrm{birth}}=\epsilon_{\mathrm{death}}$) at some time point during the simulation and extracting the time path corresponding to this pair.
\end{enumerate}
\textblue{The $L_\infty$ distance between points in persistence diagrams corresponding to consecutive time frames (Step 3) is the same distance used to calculate the bottleneck distance between two persistence diagrams (however, using the $L_2$ distance metric does not qualitatively change the output of our algorithm). The bottleneck distance is the $L_\infty$ cost of the optimal matching between points in two persistence diagrams. In practice, computational approximations are used to perform this matching. In our setting, we prioritize pairing the most significant (persistent) features through time. Given the small time step in our simulations (10 s), we expect small changes in the actin monomer unit locations between consecutive time frames, and we use the observation that such small changes in point clouds also result in small changes in the corresponding persistence diagrams (by the stability theorem in \cite{cohen2007stability}). By ordering the features from high to low persistence in Step 2 and by finding the nearest neighbor of the most persistent features in the sense of the $L_\infty$ metric in Steps 3-4, we take advantage of the intuition behind the stability theorem and expect our pairing for the most significant feature to be accurate.}
The linkage tolerance $d_\ell$ for our simulations is chosen to be $370$~nm based on the scale of the spatial domain of interest. Since in Step 1 of the algorithm we calculate all birth-death pairs, one option is to select the pair with the largest persistence and only calculate its corresponding path as outlined in step \textblue{6}, followed by steps 2-\textblue{5}. As presented, the algorithm outlines the calculation of all paths through time in the persistence diagram. Sample code is provided in the repository on \cite{Github_connectbd2020}.

The connections between points (birth-death pairs) that are close to the diagonal in the persistence diagram representation (see Figure~\ref{fig:paths}) may not necessarily correspond to the same $1$-dimensional topological hole generated from the data at later times. On the other hand, the algorithm accurately matches through time the birth-death pairs that are farthest from the diagonal in the persistence diagram, since a single persistent path corresponding to a $1$-dimensional hole emerges (see Figure~\ref{fig:bd_dvt} and Figure~\ref{fig:paths}b).

It is worth noting that the standard parameters driving the dynamics of the actin-myosin simulations considered here lead fairly consistently to at most one global emerging hole (ring) in the simulation domain, therefore the significant path extracted in step~5 of the algorithm likely corresponds to the same persistent $1$-dimensional hole in the data. Given two birth-death pairs that cross or come close together in persistence diagrams for consecutive simulation frames, our algorithm could lead to uncertainty in whether the paths extracted are properly separated.
In addition, if two birth-death pairs extracted at one time point of the simulation are closest to only one pair in the next time point, the algorithm could choose an incorrect match for a path. However, given the dynamics of the actin-myosin network in our simulations and the global $1$-dimensional hole structures we seek, this is unlikely to occur in our application. The exception to this is the matching of points close to the diagonal in the persistence diagrams, which are nonetheless below the significance threshold we discuss in the next section. 

\textblue{In the Appendix, we illustrate in Figure~\ref{fig:paths}} the results of connecting the birth-death pairs corresponding to $1$-dimensional topological hole features for consecutive times in the simulation of Figure~\ref{fig:pers_diagram_time}a and Figure~\ref{fig:bd_dvt}a. Each of the rainbow-colored paths in the persistence diagram in Figure~\ref{fig:paths}a connects through simulation time the birth-death pairs corresponding to a $1$-dimensional hole feature using our algorithm, while Figure~\ref{fig:paths}b extracts the significant path corresponding to the largest hole that emerges in the simulation domain \textblue{(see also the animation in Online Resource 1)}. \textblue{The proposed method also generalizes} to connecting birth-death pairs within a persistence diagram for higher-order cavities. The video in Online Resource 2 illustrates  both the significant path for the $1$-dimensional hole (dark orange) and all paths corresponding to the $2$-dimensional voids (blues and greens) forming through time in a standard simulation where we extract three \textblue{monomer unit} locations per filament, as shown in Figure~\ref{fig:simp_complex}b. In this example, all cavities ($2$-dimensional holes) have small persistences while a $1$-dimensional hole clearly emerges, indicating that a thin tunnel emerges from the actin-myosin network in the simulation domain.

\textblue{Motivated by investigating} the timing of ring structure formation, we consider an alternative visualization to Figure~\ref{fig:paths}b that plots the persistence $\epsilon_{\mathrm{death}}-\epsilon_{\mathrm{birth}}$, i.e. the vertical distance from each birth-death pair in the significant path to the diagonal in the persistence diagram. This visualization shows rotated persistence diagrams that are visualized in a birth-persistence coordinate system \cite{adams2017persistence,stolz2018topological} with $(\epsilon_\mathrm{birth}, \epsilon_\mathrm{death} - \epsilon_\mathrm{birth})$ coordinates. 
Figure~\ref{fig:bd_dvt}b shows the evolution of the feature persistence as a function of time in a sample simulation. While this is not the case for all applications \cite{feng2019persistent}, we will consider pairs that are farther away from the diagonal in persistence diagrams, and thus higher along the vertical axis in Figure~\ref{fig:bd_dvt}b, to correspond to $1$-dimensional hole features that are more significant. \textblue{We proceed to discussing methods for assessing feature significance in this context in \S~\ref{sec:significance}.}

\subsection{Estimating significance for topological features} \label{sec:significance}

\textblue{This work} raises the question of how we might distinguish significant features from noise in time-series data. A common method for identifying signal in persistent homology is to seek features that have a large persistence (i.e., $\epsilon_{\mathrm{death}}-\epsilon_{\mathrm{birth}}$) and therefore correspond to birth-death coordinates farther away vertically from the diagonal. This is, of course, not universally the right thing to do. Recent studies with applications to neuroscience and voting maps have addressed the limitations of this intuition since some short-lived features may hold key insights and the interpretation of feature persistence is not always clear \cite{stolz2017persistent,feng2019persistent}. In our application, large persistence is \textblue{a meaningful notion}. The point clouds we consider \textblue{are sub-sampled from the simulated} three-dimensional locations of actin \textblue{monomer units}, and Vietoris-Rips simplicial complexes are constructed based on the Euclidean pairwise distances between these points. It follows that persistent \textblue{$1$-dimensional} features (those that are farther from the diagonal \textblue{in the persistence diagram}) correspond to large-scale holes emerging in the simulation domain, for which it is our challenge to identify \textblue{whether and when they form and} how long they last (see the video in Online Resource 1).

We now describe \textblue{three methods} by which we selected the threshold for significance of feature persistence \textblue{in our application}. \textblue{The first two methods are statistical, while the third is trajectory-based and relies on the observation that there is a single emerging hole in the domain of our simulations.}
\textblue{The methods outlined below give similar estimates for the threshold level for feature persistence, which we conservatively choose to be $500$~nm. Given this threshold, we} define the onset of ring formation as the last time that the main feature's persistence crosses from below to above the significance threshold, as illustrated by the vertical and horizontal dashed lines in Figure~\ref{fig:bd_dvt}b.

\subsubsection{Significance through hypothesis testing} \label{sec:hypo_testing}

\textblue{In the hypothesis testing perspective, we need to establish an appropriate null model for randomly-distributed actin filaments. By analyzing the distribution of the maximum persistence lengths in the null model frames, we will establish a threshold so that the null model will be unlikely to generate features with persistence higher than this threshold. }

\begin{figure*}[h]
\centering
\includegraphics[width=0.9\textwidth]{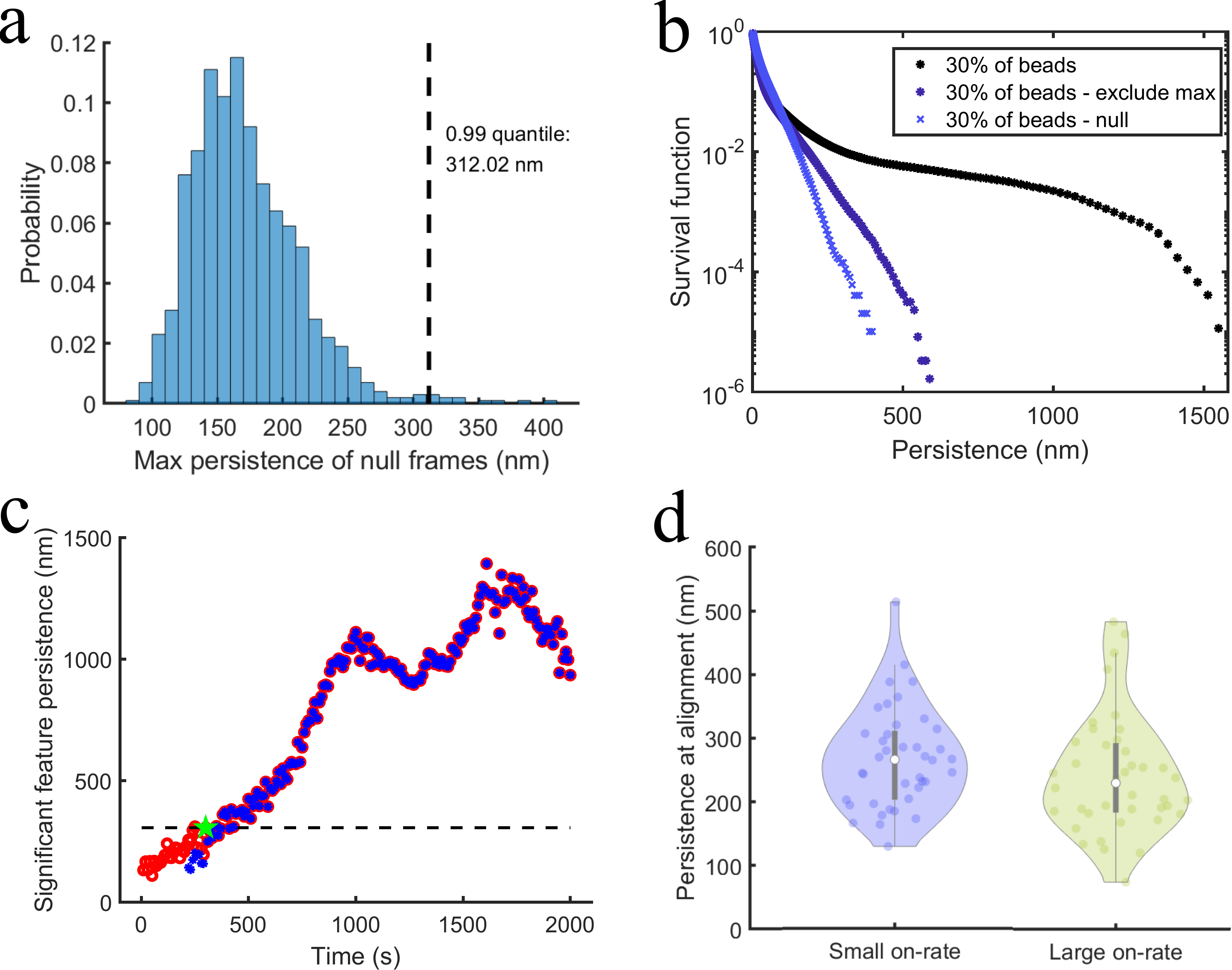}
\caption{{\bf \textblue{Assessing significance of topological features.}} 
\textblue{(a) Distribution of the maximum persistence in each generated null frame. Dashed vertical line indicates the $.99$ quantile for this distribution. (b) Survival function, i.e. proportion of features with persistence size larger than the corresponding persistence length on the $x$-axis, for the database of $35$ MEDYAN simulations (black stars), for the same simulations but excluding the maximum persistence at each time frame (purple stars), and for the null model frames (blue x's). The $y$ axis is on a log scale for ease in visualization. (c) In red, we mark the persistence level corresponding to the 1-dimensional topological feature with maximum persistence at that time. In blue, we mark the persistence corresponding to the path of the most significant hole through time. There is a time point at which the two paths align (marked by a green star). We denote the persistence level at that time point as the persistence at alignment and mark this level using the horizontal dashed line.
(d) Violin plot of the persistence level at alignment (see panel c) for 40 small on-rate simulations (blue) and 40 large on-rate simulations (green).}}
\label{fig:significance_methods}
\end{figure*}

\textblue{To develop the null model, we first record the filament lengths and the number of monomer units extracted from each filament at each time frame from the MEDYAN simulation database described in \S~\ref{sec:medyan_database}. We then generate a null frame by randomly choosing a MEDYAN-simulated frame and by constructing straight filaments with the same lengths and the same density of sampling of the actin monomer units as in the original frame. These filaments are assigned random positions (drawn uniformly at random from the simulated region) and orientations (with angle off the $x$-axis drawn uniformly from the interval $[0,2\pi]$); we reject filaments that extend outside the simulation domain (see Figure~\ref{fig:cdf}a in the Appendix for an example null frame). In this way, we generate 1000 null frames that have the same filament numbers, lengths, and the same numbers of monomer units sampled from the polymers as in model simulation frames generated in \S~\ref{sec:medyan_database}. We compute the persistent homology for the point clouds extracted from each null frame; see Figure~\ref{fig:cdf}b in the Appendix for an example of a persistence diagram corresponding to a null frame. In Figure~\ref{fig:significance_methods}a, we study the distribution of the maximum persistence from each of the $1000$ null frames and find that the $.99$ quantile of this distribution is roughly $312$~nm; this means that there is less than $1\%$ chance that a random Poisson spatial process would generate a 1-dimensional hole of this (or larger) size. Therefore, a significance threshold larger than or equal to  this level is suitable for rejecting the null hypothesis that the feature can be generated by this type of randomly drawn filament network. As we see below, this method produces a small estimate for the significance threshold, most likely because this model for noise is not rich enough to capture everything that produces spurious features in the data set. This is, in part, why we use multiple perspectives on assessing significance in this work.}

\subsubsection{Significance through analysis of spurious feature distribution} \label{sec:survival}

\textblue{In the second statistical method, we pool all the persistence lengths from all time frames in the database of MEDYAN simulations described in \S~\ref{sec:medyan_database} and compute the corresponding ``survival function'' (the probability that features have a persistence larger than a given fixed value). Figure~\ref{fig:significance_methods}b illustrates this survival function, where the $y$-axis values are shown in logarithmic scale and represent the proportion of features with persistence size larger than the corresponding persistence length on the $x$-axis. We notice the emergence of a plateau which corresponds to the transition from spurious to significant topological features of interest (marked with stars in Figure~\ref{fig:significance_methods}b). In these results, we use a sampling density of $30\%$ of the actin monomers; we further comment on the impact of the sampling density in the Appendix.} 

\textblue{We similarly compute the survival function (marked with x's in Figure~\ref{fig:significance_methods}b) corresponding to the null model frames generated as described in \S~\ref{sec:hypo_testing} (see Figure~\ref{fig:cdf}a for a sample null model frame and Figure~\ref{fig:cdf}b for its corresponding persistence diagram). The survival functions for the frames generated by the null model and the MEDYAN simulations have considerably different shapes. The emergent plateau that begins somewhere between $300$ and $500$~nm in the MEDYAN-generated frames marks a transition from spurious topological features to values that are associated with the topological features of interest. This figure shows that there are very few $1$-dimensional features with a persistence level in this interval that are generated by the random filament network in the null model (x's in Figure~\ref{fig:significance_methods}b). This is consistent with our findings in \S~\ref{sec:hypo_testing} based on the null model alone.} 

\textblue{At the suggestion of an anonymous reviewer, we also include the distribution of topological features from MEDYAN simulations if we remove the maximally persistent feature in each persistence diagram. This removal of points ensures that we are not including topological features that have emerged from the spurious feature cloud, while also including noisy effects missed in the null model. The associated survival function, marked with purple stars, shows a clear departure from the other two, but is consistent with our selection of 500~nm as a significance threshold.
}

\subsubsection{Significance through path tracing} \label{sec:alignment}

\textblue{In the third approach, we propose a means of finding the persistence length at which the most significant path (as found using the algorithm in \S~\ref{sec:algorithm}) emerges as the dominant feature in a video of actin-myosin interactions. For each model simulation in the collection described in $\S$ \ref{sec:medyan_database}, we compare the maximum persistence among all 1-dimensional features at each time to the time-dependent persistence of the significant path. In Figure~\ref{fig:significance_methods}c (corresponding to a sample simulation with standard parameters), we mark in red the maximum persistence level among all 1-dimensional topological features at each time. We also mark in blue the persistence corresponding to the path of the most significant hole through time.}

\textblue{Figure~\ref{fig:significance_methods}c shows that, after a period with noisy, short-lived loop features, these two time-dependent persistence paths align, thus providing an indicator of the threshold for significance. In Figure~\ref{fig:significance_methods}d, we visualize the persistence levels at the alignment of these paths (see green star and horizontal dashed line in Figure~\ref{fig:significance_methods}c) using violin plots for the simulation sets of small and large myosin motor binding rates described in \S~\ref{sec:medyan_small_large} (and further analyzed in \S~\ref{sec:analysis}). The mean values for these persistence levels at alignment ($267.7$~nm for small on-rate and $241.9$~nm for large on-rate) provide estimates for the significance threshold and have qualitative agreement with the threshold values predicted using the previous methods.
}

\section{Analysis of a filamentous network model}\label{sec:analysis}

To illustrate the power of the methods introduced in \S~\ref{sec:methods} in distinguishing between different dynamic behaviors, \textblue{we consider the simulation sets described in \S~\ref{sec:medyan_small_large}, where we focus on actin-myosin organization in the context of large and small on-rates (motor binding rate parameter).}

\begin{figure}[!h]
\caption{{\bf Analysis of simulations with small (a), respectively large (b) motor binding rate for two stochastic simulations.} 
Within each simulation: (Top) Distribution of actin filaments, myosin motors, and cross-linkers at the final simulation time; (Bottom) Visualization of ring emergence as time-dependent persistence of the significant path corresponding to a $1$-dimensional hole.
}
\vspace{2ex}
\includegraphics[width=\textwidth]{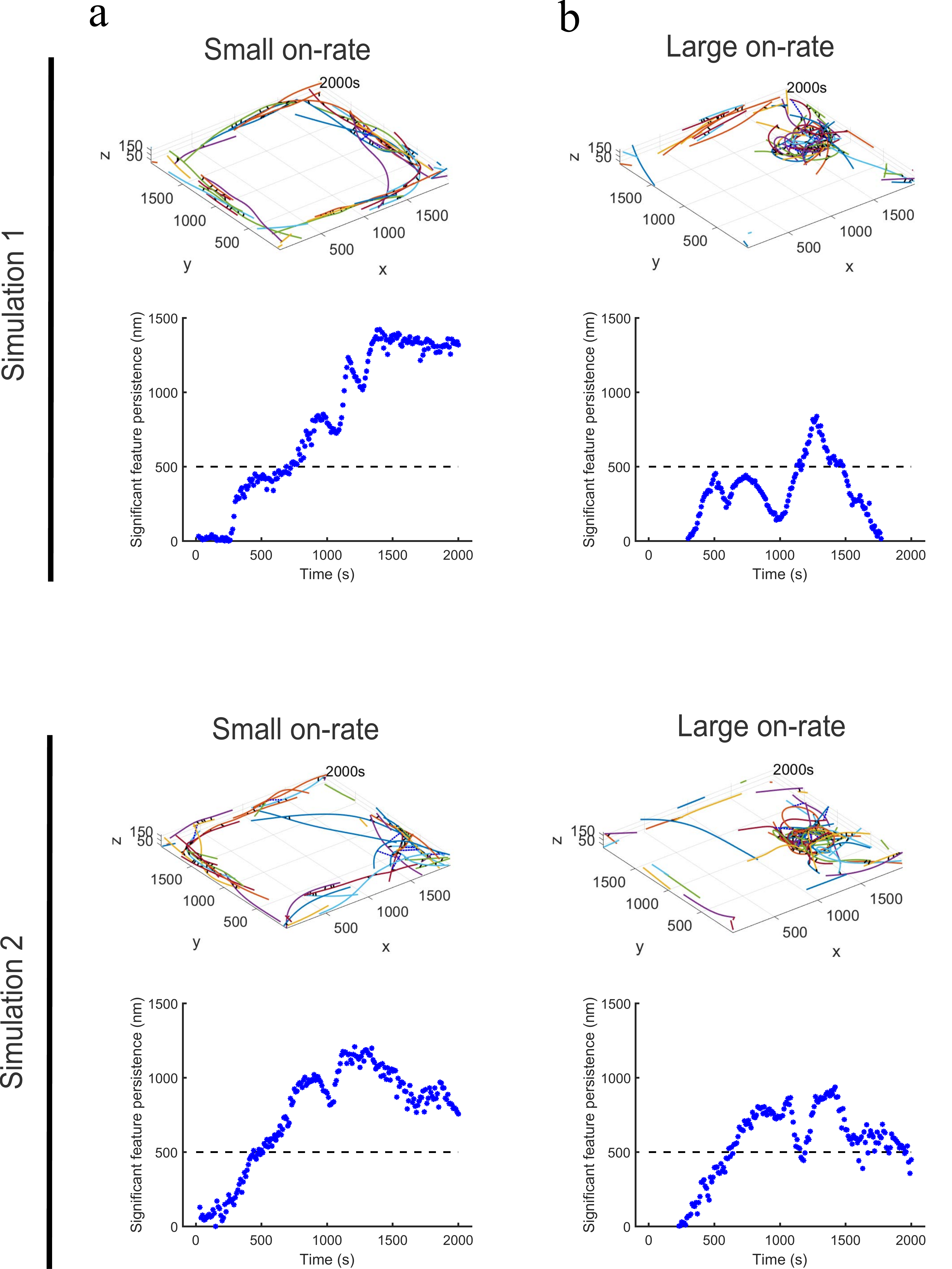}
\label{fig:small_large1}
\end{figure}

Figure~\ref{fig:small_large1} shows snapshots of the final actin configurations in two sample stochastic runs of simulations with these parameters, as well as the emergence of ring structure with time (as introduced in Figure~\ref{fig:bd_dvt}b) in each case. A decreased on-rate leads to alignment of the actin filaments at the boundaries, and thus a clear hole emerges in the simulation domain (Figure~\ref{fig:small_large1}a). A large on-rate leads to more frequent interactions between actin filaments and myosin motors and as result more contractile behavior and clustering of filaments in various regions of the domain (Figure~\ref{fig:small_large1}b).

This distinction in the dynamics is clear when visualizing the persistence $(\epsilon_\mathrm{death} - \epsilon_\mathrm{birth})$ plots in Figure~\ref{fig:small_large1}. 
The small on-rate setting corresponds to a clear hole in the simulation domain that maintains high feature persistence value (i.e., it is far from the diagonal in the persistence diagram) as time progresses in each of the two simulations illustrated. In the large on-rate case, a hole forms but is not maintained over time and therefore the significant feature does not persist throughout the simulation. When the myosin motors have a higher likelihood of binding to actin filaments, the dynamics of the polymer network shows more variability, as illustrated by outcomes from application of our technique to two stochastic realizations of the dynamics with this parameter choice in Figure~\ref{fig:small_large1}b. Here the significant path captures several short-lived holes that are not maintained throughout time.

\begin{figure}[!h]
\caption{{\bf \textblue{Histograms} of the time and persistence (size) of \textblue{the first} significant ring \textblue{emergence} in small vs. large on-rate simulations.} \textblue{(a) Histograms of the time of \textblue{the first} significant ring onset in blue (small on-rate) and in green (large on-rate) based on 40 simulations each. (b) Histograms of the ring size (maximum persistence) in blue (small on-rate) and in green (large on-rate) based on 40 simulations each.}}
\centering
\includegraphics[width=0.9\textwidth]{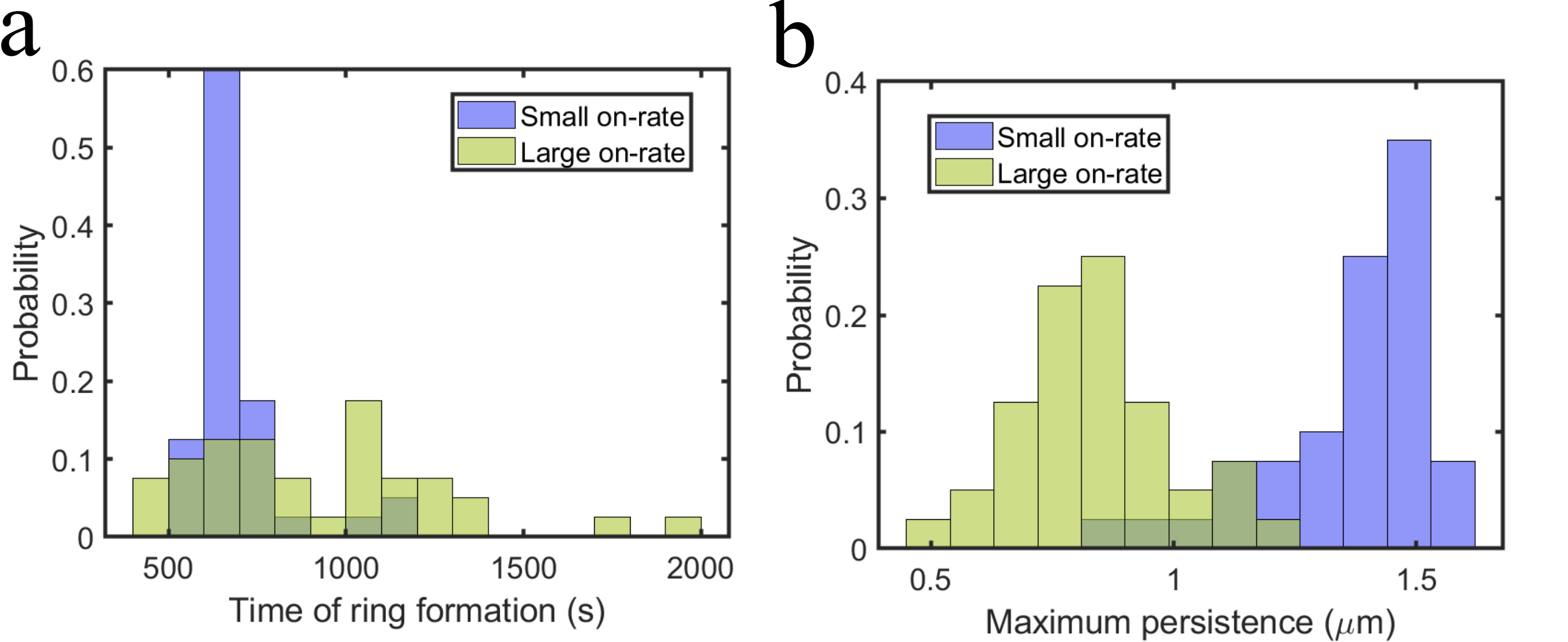}
\label{fig:dens_estimation}
\end{figure}

Our algorithm allows us not only to observe these distinguishing features, but also to quantify them in the \textblue{40} MEDYAN simulations of the actin-myosin interactions for each setting: half on-rate and double on-rate \textblue{(in reference to the standard myosin-2 parameters in \cite{popov2016medyan})}. For each parameter choice, we are interested in exploring patterns related to the number of emergent holes, whether they close or remain open, the lifespan of the significant holes, and their size -- quantified as the maximum persistence of the topological feature during the time it is significant. \textblue{As outlined in \S~\ref{sec:significance}, we use a conservative significance level threshold of $500$~nm for this analysis, as validated by the three methods for establishing significance. When carrying out the same analysis for a threshold of $300$~nm, the qualitative conclusions were the same: the small on-rate simulations produced larger holes more reliably than the large on-rate simulations.}

In the small on-rate simulations, the dynamics is consistently characterized by alignment of filaments at the domain boundaries (see Figure~\ref{fig:small_large1}a). In all runs, our method identifies a significant hole that persists until the end of the simulation time, allowing us to quantify the ensemble statistics of the timing of ring formation. The estimate for the mean time of ring formation is \textblue{$693\pm 61.81$~s} (this and all the following intervals are at the $95\%$ confidence level). We illustrate the distribution of the timing of significant ring formation in \textblue{the histogram in} Figure~\ref{fig:dens_estimation}a in blue. We can describe the size of the detected rings by the maximum persistence ($\epsilon_\mathrm{death} - \epsilon_\mathrm{birth}$ in the persistence diagram) over the time that the ring is above the significance level. Figure~\ref{fig:dens_estimation}b illustrates the histogram of the ring sizes in blue, where the estimate \textblue{for the mean} of the maximum persistence size achieved by significant holes in these 40 simulations is \textblue{$1.36 \pm 0.07~\mu$m}.

As illustrated in Figure~\ref{fig:small_large1}b, the large on-rate simulations lead to more contractility of the actin-myosin network and to a higher likelihood of polymer clusters forming in the domain. The estimate for the \textblue{mean} number of significant holes emerging throughout these simulations is \textblue{$4.02 \pm 1.22$} (see Simulation 2 in Figure~\ref{fig:small_large1}b for an example of a run that yields several short-lived significant holes). The holes in the actin-myosin network identified by our method for these simulations are on average smaller and shorter-lived than those in the small on-rate simulations, with an estimate for the mean hole lifespan of \textblue{$226.21 \pm 88.82$~s} and for the \textblue{mean of the} maximum persistence of \textblue{$0.85 \pm 0.07$~$\mu$m}. Figure~\ref{fig:dens_estimation}a shows the \textblue{histogram of the} distribution for the time of ring formation for the first hole emerging in the large on-rate simulations in green. In some stochastic runs, the onset time is similar to that of the significant holes in the small on-rate simulations, however the holes in the large on-rate simulations fall apart as their persistence soon goes below the significance threshold (see Figure~\ref{fig:small_large1}b for two representative examples). In other stochastic runs, a significant ring does not emerge until much later in the simulation\textblue{, so that the variance of the time of significant hole formation is considerably larger for the higher binding rate setting.} In Figure~\ref{fig:dens_estimation}b, we illustrate \textblue{the histogram of} the maximum persistence over all rings in each large on-rate simulation in green. As expected given that the rings are less prominent in the large on-rate simulations, this distribution is shifted to the left (corresponding to smaller maximum ring persistence) compared to the small on-rate simulations.

Therefore, our proposed approach for analyzing time-series data of cellular interactions is able to rigorously distinguish between and quantify emerging features of parameter changes in agent-based model simulations of cellular polymer interactions.

\section{Conclusions}
\label{sec:conclusions}
Understanding complicated interactions of filamentous networks and multiple chemical species at the cellular level often requires complex simulations that provide insight into the temporal and spatial dynamics of the interacting proteins. Here we carried out numerical simulations of actin-myosin and crosslinker interactions using the MEDYAN model \cite{popov2016medyan}. Filament contractility and alignment in these models have been studied using classical tools such as calculation of the network radius of gyration and of an orientational order parameter of the system \cite{popov2016medyan}. However, an understanding of how filamentous networks interact to create higher-order structure and organization in cells is lacking. We propose a computational technique based on topological data analysis to identify ring structure in complex simulations of filament organization.

The method we propose requires that we sample the spatial distribution of filaments at each time point of a dynamical simulation and thus extract a point cloud. Computing the \textblue{persistent homology} of this discrete set of data points generates a persistence diagram, which is commonly used to represent and visualize birth-death pairs corresponding to topological objects such as loops. \textblue{We present an algorithm that connects significant pairs across multiple persistence diagrams over time. As previously mentioned, \cite{cohen2006vines} introduced this concept, called vines and vineyards, defined as continuous families of persistence diagrams for time series of continuous functions.} Computation of these vineyards requires a list of simplices at each time point \textblue{(and relies on sublevel set filtrations)}, while our algorithm only requires knowledge of the persistence diagrams (birth-death pairs) at each time. Taking advantage of the relative simplicity of the dominant topological feature we pursue, our proposed approach has the advantage that it only requires topological summaries of the data. \textblue{We also present multiple perspectives on significance of features in persistence diagrams. We emphasize transparent computational methods for detecting and quantifying the most significant higher-order structure that emerges from polymer network interactions and evolves in time according to a stochastic dynamical system. Individually, these perspectives provide an incomplete view of the persistence distribution of spurious topological features. Taken together, however, the resonance among these perspectives provides a robust assessment of when the dominant feature of the dynamics emerges and how long it endures.}

\section*{Acknowledgments}
MVC was supported by The Ohio State University President's Postdoctoral Scholars Program and by the Mathematical Biosciences Institute at The Ohio State University through NSF DMS-1440386.
ATD is supported by NSF DMS-1554896. RJ and SAM are supported by NIH R01GM122082-01. 

Part of this research was conducted using computational resources and services at the Ohio Supercomputer Center through the Mathematical Biosciences Institute at The Ohio State University. We are grateful to Dr. Chad Topaz, Dr. Henry Adams, Dr. Sarah Day, Dr. Peter Bubenik, \textblue{Dr. Brittany Fasy and her group}, and Dr. Garegin Papoian and his lab for helpful conversations in developing this work. 

\textblue{We also thank the anonymous referee who suggested carrying out a rigorous hypothesis test for establishing feature significance.}

\section*{Appendix}

\subsection*{\textbf{A1.} \textblue{Visualization of birth-death pairs connected as paths through time in the persistence diagram}}

\begin{figure}[!h]
\caption{{\bf Visualization of pairs of birth-death proximity as paths connected through time in the persistence diagram.}
(a) Each rainbow-colored path corresponds to a topological hole identified and connected through time using our algorithm. (b) The most significant path is isolated. Note that these paths are only meaningful outside a noisy threshold above the diagonal that we discuss in \textblue{\S~\ref{sec:significance} on estimating a significance threshold}.}
\vspace{2ex}
\includegraphics[width=0.9\textwidth]{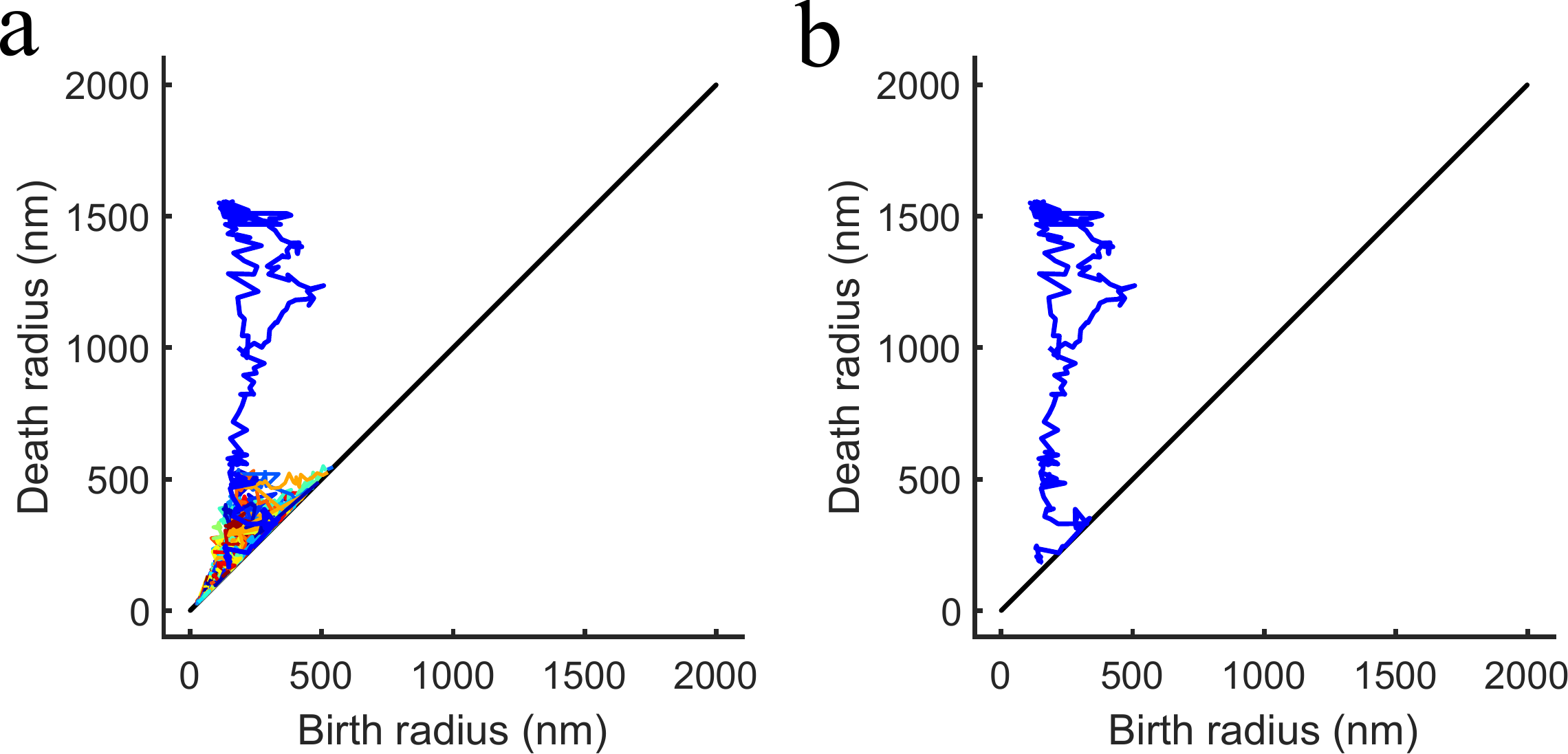}
\label{fig:paths}
\end{figure}

\textblue{One approach to illustrating the results of our algorithm on a sample MEDYAN simulation is shown in Figure~\ref{fig:paths}, where we connect the birth-death pairs corresponding to $1$-dimensional topological hole features at consecutive times in the simulation.} The points $(\epsilon_{\mathrm{birth}},\epsilon_{\mathrm{death}})$ (marked by red triangles in Figure~\ref{fig:pers_diagram_time}c) are connected with straight line segments \textblue{in Figure~\ref{fig:paths}}. The algorithm described \textblue{in \S~\ref{sec:algorithm}} ensures that these segments \textblue{(which connect the features through time)} are relatively short for each path. \textblue{In Figure~\ref{fig:paths}b,} we isolate the significant path corresponding to the largest hole that emerges in the simulation domain. This is achieved by identifying the path that has the largest $\epsilon_{\mathrm{death}}-\epsilon_{\mathrm{birth}}$ persistence at some time point in the simulation. We visualize this significant path separately in Figure~\ref{fig:paths}b. The formation of this path and its departure away from the diagonal (i.e., the addition of birth-death pairs at subsequent times) are also illustrated in the animation in Online Resource 1.

\subsection*{\textbf{A2.} Effect of \textblue{monomer unit} sampling on significant paths}

\begin{figure*}[h]
\centering
\includegraphics[width=0.9\textwidth]{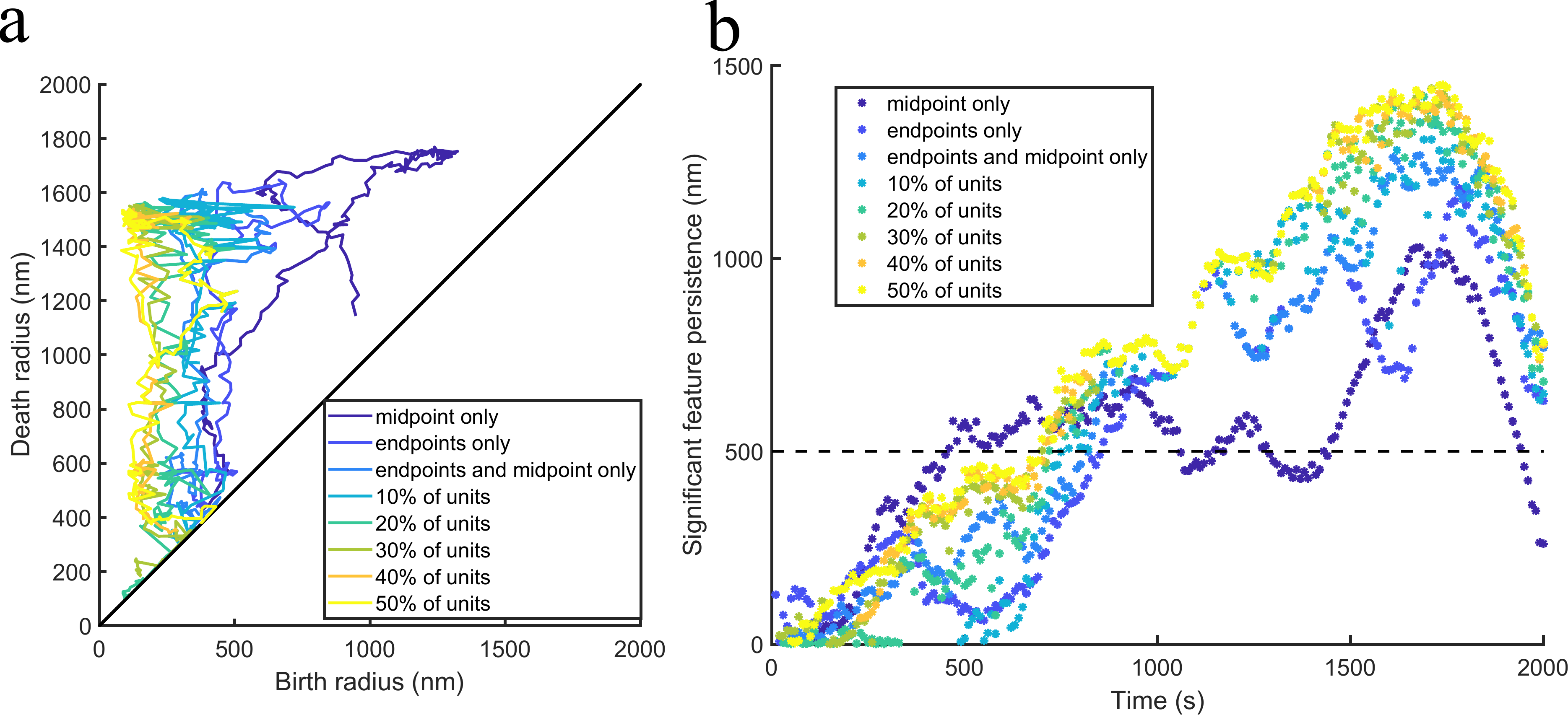}
\caption{{\bf \textblue{Sensitivity to the sampling density of the actin filaments.}}(a) Visualization of the most significant topological hole path within a persistence diagram for an increasing fraction of actin \textblue{monomer units} sampled along each filament in a standard binding rate MEDYAN simulation. (b) Visualization of the time-dependent persistence for an increasing fraction of actin \textblue{units} sampled along each filament.}
\label{fig:sampling_density}
\end{figure*}

We explore how the significant paths corresponding to $1$-dimensional features (as in Figure~\ref{fig:paths}) and the time-dependent persistence plots (as in Figure~\ref{fig:bd_dvt}b) change with an increasing fraction of monomer units extracted from each simulated filament. Figure~\ref{fig:sampling_density}a shows that sampling more units (points) along the filaments leads to the significant paths moving left in the persistence diagram but converging at a mid-range of the sampling density. We expect this to be the case since the more units we extract, the closer the points are in the point cloud, and therefore we observe the $1$-dimensional holes at a smaller birth radius. As the percentage of monomer units extracted along each filament increases, we also observe that the paths are more persistent over time in Figure~\ref{fig:sampling_density}b. The onset of the significant hole and its maximum persistence are similar for mid-range to high sampling densities. On the other hand, only extracting the midpoint of the actin filaments leads to a significant path that is not above the persistence threshold consistently through time. As expected for persistent homology calculations, considering large point clouds extracted from a larger fraction of the monomer units along each filament increases the computational time of the algorithm proposed. 

\subsection*{\textbf{A3.} \textblue{Effect of monomer unit sampling on significance through spurious feature distribution}}

\begin{figure*}[h]
\centering
\includegraphics[width=0.9\textwidth]{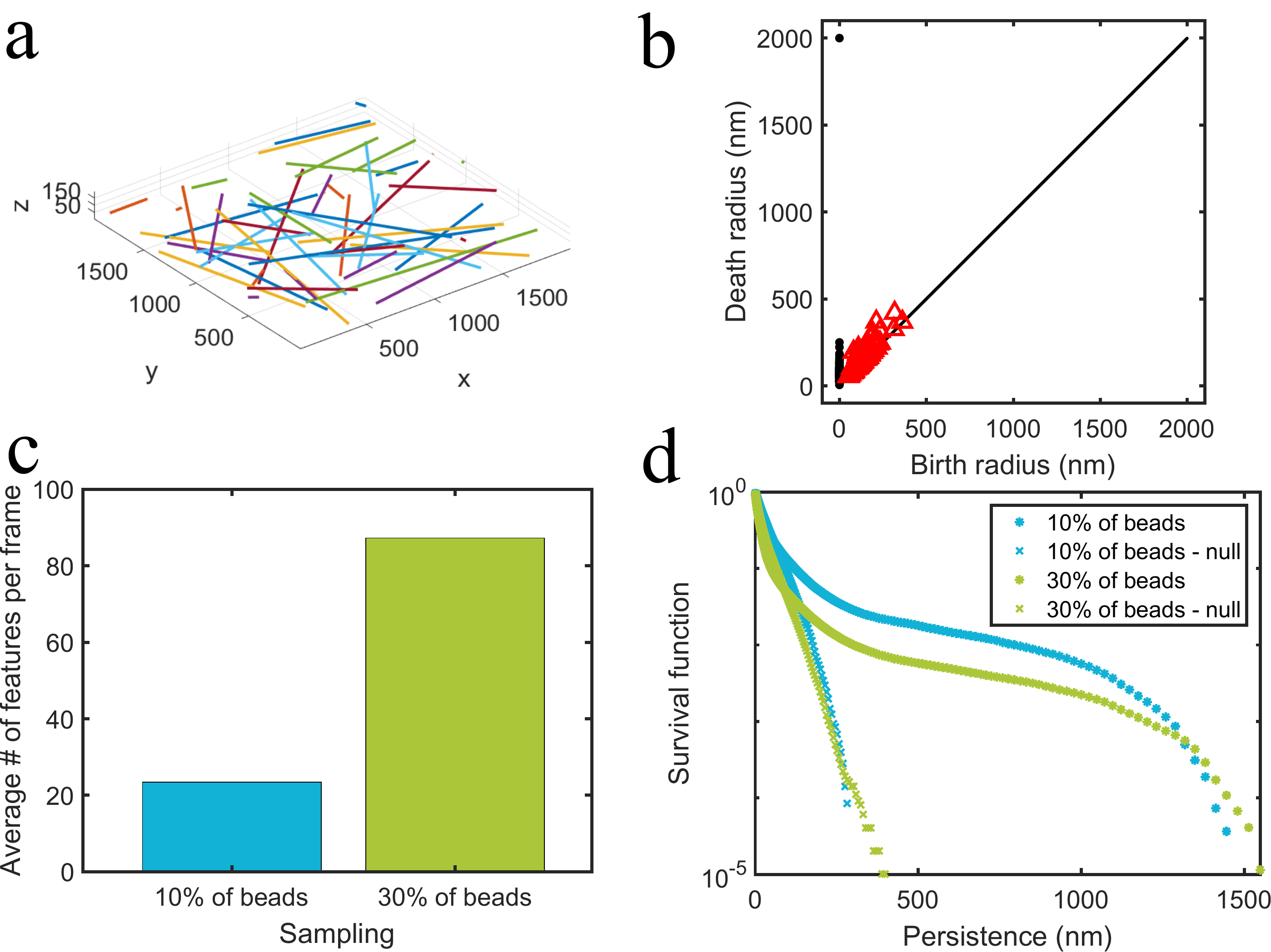}
\caption{{\bf \textblue{Assessing significance through the distribution of persistence and comparison with the null model.}} \textblue{(a) Sample frame of actin filaments generated according to the null model described in the main text. (b) The corresponding persistence diagram for the null model frame in (a); black circles correspond to connected components and red triangles correspond to loops.} (c) Average number of $1$-dimensional features per time frame for two sampling densities of the actin \textblue{monomer units} in our database of $35$ MEDYAN simulations. (d) Survival function, i.e. proportion of features that are larger than the corresponding persistence length on the $x$-axis for the \textblue{database of $35$ MEDYAN simulations (stars) and for the null model frames (x's)}. The $y$ axis is on a log scale for ease in visualization.}
\label{fig:cdf}
\end{figure*}

\textblue{Figure~\ref{fig:cdf}a shows an example of a null frame generated using the framework described in \S~\ref{sec:hypo_testing}, where actin filaments are assigned random locations and orientations in a typical simulation domain. In Figure~\ref{fig:cdf}b, we illustrate the corresponding persistence diagram for the point cloud of monomer units extracted from the frame in Figure~\ref{fig:cdf}a. Here we focus on the impact of the sampling density of actin monomer units on our method for establishing significance as described in \S~\ref{sec:survival}. We focus on two sampling densities used to generate Figure \ref{fig:sampling_density}, which produced similar results for the time-dependent paths ($10\%$ and $30\%$). In Figure~\ref{fig:cdf}c, we compute the average number of topological features in each video frame. The $30\%$ sample density produced approximately four times the number of features produced using the $10\%$ sampling. While we expect to observe more $1$-dimensional features in the larger point clouds corresponding to the $30\%$ sampling density, we also lack a general understanding of how many noise-induced features will appear for a given simulation or for different sampling densities.  
}

\textblue{In Figure~\ref{fig:cdf}d, we study the distribution of spurious topological features for both null model frames and MEDYAN model simulation frames for the two sampling densities. We recall from \S~\ref{sec:survival} that, to generate Figure~\ref{fig:cdf}d, we pool all persistence lengths for each sampling density and compute the ``survival functions'' corresponding to the null model frames (x's, see Figure~\ref{fig:cdf}a for a sample null frame) and corresponding to the model-generated frames (stars, see Figure~\ref{fig:pers_diagram_time}a for sample simulation frames). As observed in the text, and for both sampling densities, the survival functions for the null and model-generated frames have significantly different shapes.}

\textblue{When comparing the survival function plots for the model-generated frames using the two sampling densities (marked with stars in Figure~\ref{fig:cdf}d), we note that while the tail behavior and the proportion values differ, the emergent plateau that begins between 300 and 500~nm is present in both curves. This shows that the choice of 500~nm for the significance threshold is not dependent on the sampling density of the actin filaments. }

\bibliographystyle{unsrt}  

\bibliography{references} 

\end{document}